\documentclass[summary]{URSIRCRS2020}

\title{Assessment of the Performance of Ionospheric Models with NavIC Observations during Geomagnetic Storms}

\author{Sumanjit Chakraborty*\affref{ref1} and Abhirup Datta\affref{ref1}}

\affiliation{
\aff{ref1}{Discipline of Astronomy, Astrophysics and Space Engineering, Indian Institute of Technology Indore. Simrol Campus. Indore- 453552, Madhya Pradesh, India}

}

\begin{document}

\maketitle

\begin{abstract}
The paper presents an assessment of the performances of the global empirical models: International Reference Ionosphere (IRI)-2016 and the NeQuick2 model derived ionospheric Total Electron Content (TEC) with respect to the Navigation with Indian Constellation (NavIC)/ Indian Regional Navigation Satellite System(IRNSS) estimated TEC under geomagnetic storm conditions. The present study is carried out over Indore (Geographic: 22.52$^{\circ}$N 75.92$^{\circ}$E and Magnetic Dip: 32.23$^{\circ}$N, located close to the northern crest of the Equatorial Ionization Anomaly (EIA) region of the Indian sector). Analysis has been performed for an intense storm (September 6-10, 2017), a moderate storm (September 26-30, 2017) and a mild storm (January 17-21, 2018) that fall in the declining phase of the present solar cycle. It is observed that both IRI-2016 and NeQuick2 derived TEC are underestimates when compared with the observed TEC from NavIC and therefore fail to predict storm time changes in TEC over this region and requires real data inclusion from NavIC for better prediction over the variable Indian longitude sector. 
\end{abstract}

\section{Introduction}

The low latitude ionosphere consists of several features, such as the equatorial ionization anomaly (EIA), equatorial electrojet, equatorial plasma fountain, spread F and plasma bubbles, as a result of the horizontal orientation of the geomagnetic field at the geomagnetic equator [2]. It is expected that as the sun shines over the geographic equator, the ion and electron density should be maximum around that region and will go on decreasing towards the poles. Measured values show that this density has peculiar crests around $\pm$ 15$^{\circ}$ magnetic latitude and trough around the magnetic equator [1]. The ionosphere over central India falls under this anomaly region where sharp latitudinal gradient in the ionization is observed. The latitudes which fall in the EIA has the highest concentration of electron density and is nearly about 70\% of the global density distribution. The ionospheric total electron content (TEC) is a vital parameter of the ionosphere and is defined as total number of electrons integrated between two points, along a tube of unit cross sectional area and is expressed in TECU, where 1 TECU = 10$^{16}$ electrons/m$^2$. It gets enhanced or depleted as a result of positive or negative geomagnetic storms.   

Geomagnetic storms are temporary disturbances of the magnetosphere of earth. They are caused by the solar wind shock wave which interacts with the geomagnetic field. The increase in the solar wind pressure initially compresses the magnetosphere [5]. Whenever there are periods of such magnetic disturbance, the horizontal component of the Earth's magnetic field (H) gets depressed. The recovery to its average value is gradual. Earlier studies show that at the mid-latitudes and the latitudes at the equatorial region, the decrease in H can be represented by a uniform magnetic field parallel to the geomagnetic dipole axis and that it is directed southward. The magnitude of this disturbance field which is axially symmetric in nature, varies with the storm-time or the time measured from the onset of the storm. This onset can be understood to be as a sudden increase in the value of H globally, this is well known in literature as the storm sudden commencement (SSC). Following this SSC, H remains above its average level for a few hours, this is known as the initial phase of the storm. It is followed by a very large decrease in H which is globally observed and it indicates the main phase of the storm. The magnitude of this decrease in H indicates how severe the storm is and the variation changes from storm to storm. The disturbance field which is represented by Disturbance storm time (Dst) index, is symmetric axially with respect to the dipole axis (http://wdc.kugi.kyoto-u.ac.jp/dstdir/dst2/onDstindex.html). Severity of geomagnetic storms can be classified [8] as: 
Dst $>$-50 nT signifying a mild storm; -50 nT $\leq$ Dst $<$ -100 nT signifying a moderate storm and -100 nT $\leq$ Dst $<$ -200 nT signifying an intense storm. As a result it is essential to study model performances during disturbed ionospheric conditions due to the geomagnetic storms in order to verify the prediction potentials of such models.

The ionosphere over Indore (22.52$^{\circ}$N and 75.92$^{\circ}$E geographic; magnetic dip: 32.23$^{\circ}$N) falls near the anomaly crest in the Indian longitude sector. In this paper, for the first time to the best of our knowledge, results of the ionospheric model derived TEC from the IRI and the NeQuick models have been compared with the NavIC estimated TEC over Indore, under intense, moderate and mild geomagnetic storms during 2017 and 2018, falling in the declining phase of the present solar cycle.

\section{Ionospheric Models and Observed Data}

The International Reference Ionosphere (IRI) is an empirical model of the ionosphere. The sources of data to this model are the incoherent scatter radars and the dense worldwide network of ionosondes along with the Alouette topside sounders in-situ instruments on board satellites. The model output provides the electron temperature and density, ion temperature and composition and the TEC from 50 km to 2000 km altitude range [13].

NeQuick2 model is an upgraded version of the NeQuick model. This model uses, a modified DGR profile formulation [4] that consists five semi-Epstein layers [12] with modelled thickness parameters [11], for describing the ionospheric electron density from 90 km to peak of F2 layer. The model topside is represented by a semi-Epstein layer with a height-dependent thickness parameter that is determined empirically [6,3]. The inputs to this model are the position(latitude, longitude) and either solar flux or sunspot number. Specific routines are present in NeQuick to evaluate electron density and the corresponding TEC by the method of numerical integration [10].

Navigation with Indian Constellation (NavIC), a regional satellite navigation system developed by ISRO, has a space segment which consists of Geostationary Earth Orbit(GEO) and Geosynchronous Orbit(GSO) satellites. The primary target of developing the NavIC is to provide information on positional accuracy not only to the Indian users but also to regions of 1500 km from its boundary, designated as its primary service area. It also has provisions for an extended service area that lies between the primary service area and area enclosed by the rectangular grid having latitudinal extent of 30$^\circ$S to 50$^\circ$N and longitudinal extent of 30$^\circ$E to 130$^\circ$E. Three of the satellites are GEO while the remaining are GSO. The sub-satellite positions of the satellites are such that all of them have continuous radio visibility with the Indian control stations. The GSO have an orbital inclination of 29$^\circ$. These satellites broadcast signals in 24 MHz bandwidth of spectrum in the L5 and S band having carrier frequencies 1176.45 and 2492.03 MHz respectively [9].

\section{Methodology}

A NavIC receiver, provided by the Space Applications Centre (SAC), ISRO, capable of receiving NavIC L5 and S1 signals along with GPS L1 signal, is operational in the Discipline of Astronomy, Astrophysics and Space Engineering, Indian Institute of Technology, Indore. An elevation angle higher than 20$^\circ$ has been chosen for the NavIC values in order to avoid multipath error. The receiver provides the iono-delay at its output with a 1 Hz resolution. This iono-delay is converted to the slant TEC (STEC) [7] by the formula:
\begin{equation} 
\nu = \frac{40.3}{f^2} . {TEC}
\end{equation} 
where \textit{v} is the iono-delay, \textit{f} is the operational frequency of the signal emitted by satellites in Hz. This sTEC is converted to the equivalent vertical TEC (VTEC) [9] by the mapping factor: 
\begin{equation}
           M(E)  = \Bigg[ \bigg[ 1 -  \Big[\frac{{R_e} . cos{(E)}}{R_e+h_I}\Big]^{2}\bigg]\Bigg]^{-1/2}
\end{equation}
where {$R_e$} is the radius of the Earth (6371 km), {$h_I$} denotes the altitude of the thin shell model of the ionosphere (350 km) and ({$E$}) is the elevation angle of the space vehicle.

\section{Results and Discussions}

In this section, the diurnal variations of VTEC from the IRI and NeQuick models have been compared with the NavIC estimated VTEC and the deviations on the disturbed days from the selected periods of the geomagnetic storms are presented. 

Figure 1 shows the variation of Dst(nT) as a function of UT (h) for the 3 storms selected based on their severity. The intense storm period (September 6-10, 2017) is depicted by red, the moderate storm (September 26-30, 2017) by green and the mild storm (January 17-21, 2018) by blue. The middle day (48-72 UT) out of 5 days for all the three storms had been the disturbed day with Dst reaching the minimum. Table 1 shows the minimum Dst (nT) values for the selected storms along with the corresponding time at which minimum Dst values were observed.
\begin{figure}[ht!]
\includegraphics[width=\columnwidth,height=3.25in]{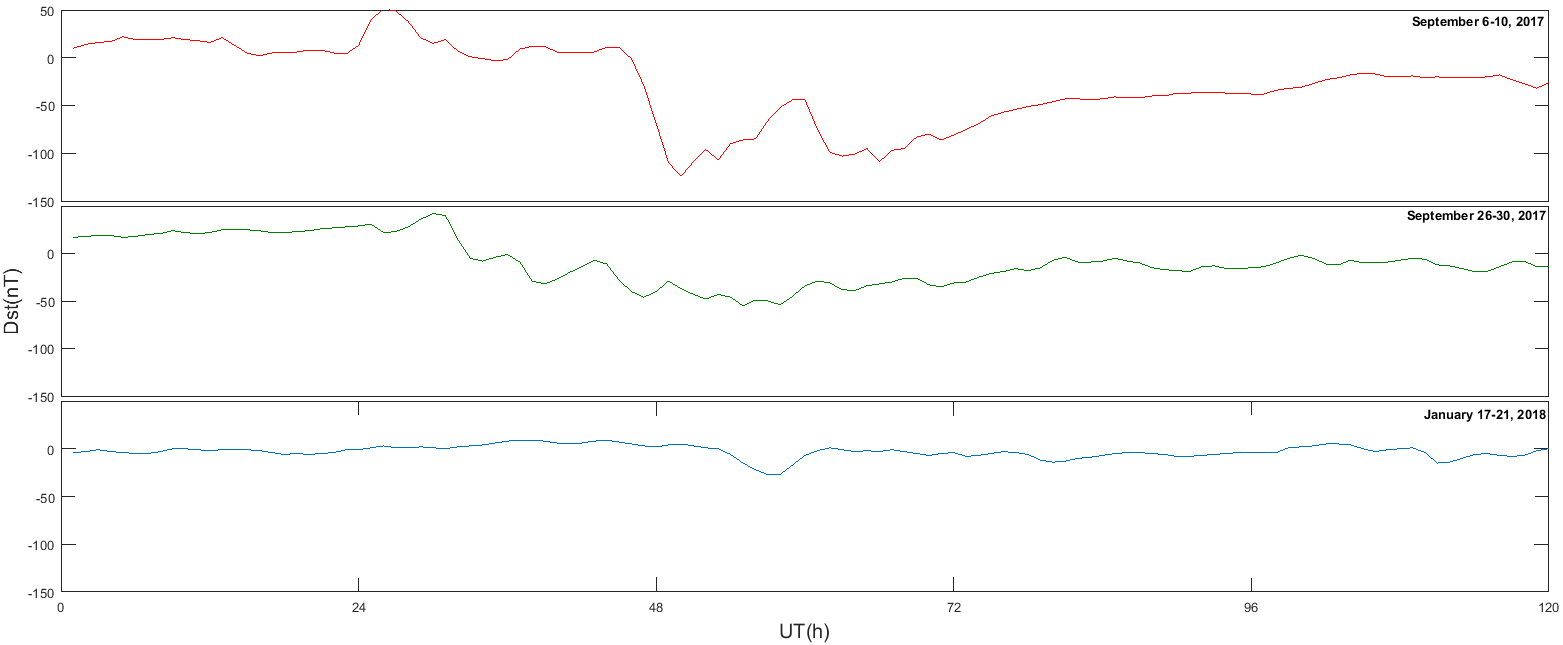}\caption{Variation of Dst(nT) with UT(h) for the intense storm of September 8, 2017(red), moderate storm of September 28, 2017(green) and mild storm of January 19, 2018(blue). The Dst dropping below -100 nT on September 8 in the top panel signifies intense, while Dst values in the middle and bottom panel on September 28 and January 19 signify moderate and mild storm respectively.}
\end{figure}
\begin{table}[ht!]
\caption{Dst values for the selected storms}
\begin{center}
\begin{tabular}{|l|c|c|c|}
\hline
\textbf{Date} & \textbf{\textit{Time (UT)}}& \textbf{\textit{Minimum Dst (nT)}} \\
\hline
September 08, 2017 & 02:00 & -124 \\
September 28, 2017 & 07:00 & -55  \\
January   19, 2018 & 09:00 & -27  \\
\hline
\end{tabular}
\end{center}
\end{table}

The diurnal variations IRI-2016 and NeQuick2 model derived VTEC are compared with the NavIC estimated VTEC during the storm period of September 6-10, 2017 is depicted in Figure 2. It is observed that although NavIC shows enhancements in VTEC on September 7 and 8, there is no variation at all in IRI while the values go on decreasing from September 6 onward in the NeQuick.
\begin{figure}[ht!]
\includegraphics[width=\columnwidth,height=3.25in]{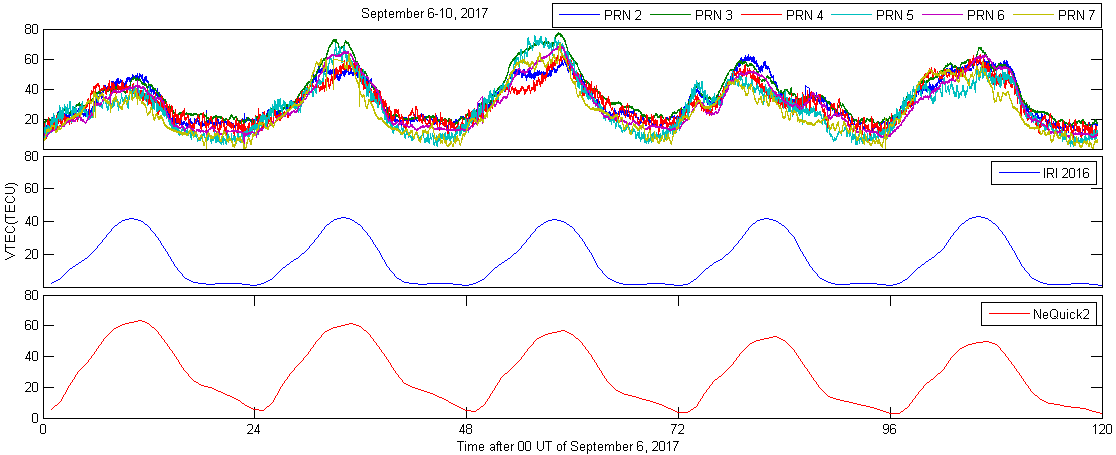}
\caption{Diurnal variation of NavIC estimated VTEC along with IRI and NeQuick derived VTEC during September 6-10, 2017}
\end{figure}

Similarly, Figures 3 and 4 show diurnal VTEC variation for the moderate and the mild storms of September 28, 2017 and January 19, 2018 respectively. In Figure 3 NavIC estimated values show enhancements on September 28,29 and 30 which is not at all captured by both the models while in Figure 4, higher VTEC values observed from NavIC on January 17, 2018 is not captured by these models. This suggests the poor prediction capability of the two models during moderate to quiet conditions of the ionosphere. 
\begin{figure}[ht!]
\includegraphics[width=\columnwidth,height=3.25in]{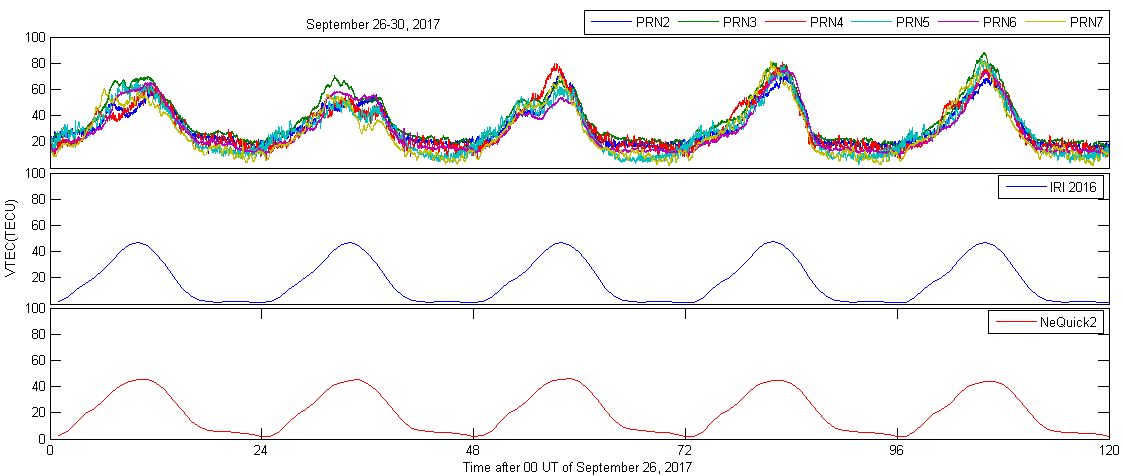}
\caption{Diurnal variation of NavIC estimated VTEC along with IRI and NeQuick derived VTEC during September 26-30, 2017}
\end{figure}

\begin{figure}[ht!]
\includegraphics[width=\columnwidth,height=3.25in]{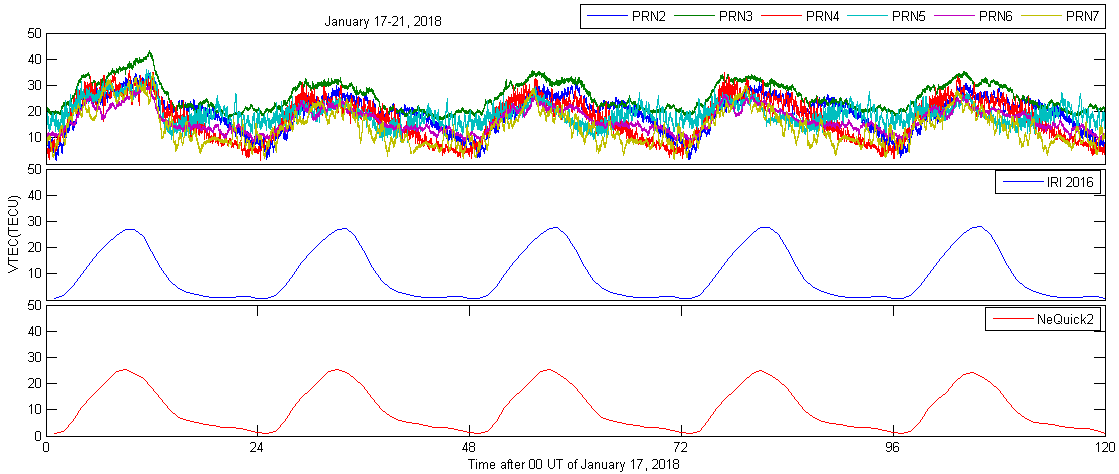}
\caption{Diurnal variation of NavIC estimated VTEC along with IRI and NeQuick derived VTEC during January 17-21, 2018}
\end{figure}

Finally, on the days when Dst dropped to a minimum from all the 3 storms, the diurnal maximum VTEC obtained from the two models and the observed VTEC from NavIC are summarised in Table 2.
Table 3 shows the deviations of model derived values with the observed ones under varying storm conditions. For the intense storm, IRI values show higher deviation from the observed VTEC while NeQuick presents greater deviations during the moderate and the mild storms. 
It can be observed that for all the three storms NeQuick and IRI derived VTEC underestimates the observed NavIC VTEC and the deviation from NavIC observed values is highest during the intense storm.The study points out to the fact that during disturbed ionospheric conditions, both IRI-2016 and NeQuick2 derived TEC are not very reliable and require modifications for a more realistic predictions in and around the anomaly region. Inclusion of real time data from the NavIC receivers could help in better prediction over the Indian longitude sector.
\begin{table}[ht!] 
\caption{Disturbed Day Peak TEC from NeQuick2, IRI-2016 and NavIC}
\begin{center}
\begin{tabular}{|l|c|c|c|c|}
\hline
\textbf{Day} & \textbf{\textit{NeQuick-TEC}}& \textbf{\textit{IRI-TEC}} & \textbf{\textit{NavIC-TEC}}\\
\hline
Sep 08, 2017 & 55.80 & 43.60 & 77.43 \\
Sep 28, 2017 & 45.72 & 47.80 & 79.13 \\
Jan 19, 2018 & 23.99 & 28.20 & 35.63\\
\hline
\end{tabular}
\end{center}
\end{table}

\begin{table}[ht!]
\caption{Deviation between Observed and Model derived TEC}
\begin{center}
\begin{tabular}{|l|c|c|c}
\hline
\textbf{Day} & \textbf{\textit{NeQuick-TEC}}& \textbf{\textit{IRI-TEC}}\\
\hline
Sep 08, 2017 & 21.63 & 33.83 \\
Sep 28, 2017 & 33.41 & 31.33 \\
Jan 19, 2018 & 11.63 &  7.43 \\
\hline
\end{tabular}
\end{center}
\end{table}

\section{Conclusions}

The paper, for the first time to the best of our knowledge, presents an analysis of the performances of the empirical ionospheric models IRI 2016 and NeQuick during the geomagnetic storms of varying severity over Indore, located near to the ionization anomaly crest. It is observed that for all the three storms, the models are unable to capture the enhancement or depletion of ionization caused as a result of the geomagnetic storms. The deviations are of higher magnitude during the intense and moderate storm while it is comparatively lower for the mild storm. These observations signify that the prediction capabilities of these models under storm conditions are not reliable and therefore require modifications and inclusion of data from NavIC, which is conceived for accurate analysis of the ionosphere, to the existing database in order deliver a more realistic values and lesser deviations from observations of the ionosphere near an anomaly region like Indore.      

\section{Acknowledgements}

SC acknowledges Space Applications Centre (SAC), ISRO for providing fellowship to pursue his research. The authors acknowledge SAC, ISRO for providing the NavIC data (ACCORD receiver) under the project number: NGP-17 to the Discipline of Astronomy, Astrophysics and Space Engineering, IIT Indore. 
Further acknowledgements go to Ms. Deepthi Ayyagari and Dr. Saurabh Das for helpful discussions.

\end{document}